**Assessment of a nonempirical semilocal density functional on solids and surfaces**

Yuxiang Mo[1,2], Roberto Car[3], Viktor N. Staroverov[4], Gustavo E. Scuseria[5], and Jianmin Tao[1,2,*]

[1]Department of Physics, Temple University, Philadelphia, PA 19122-1801, USA
[2]Department of Chemistry, University of Pennsylvania, Philadelphia, PA 19104-6323, USA
[3]Department of Chemistry, Princeton University, Princeton, NJ 08544, USA
[4]Department of Chemistry, The University of Western Ontario, London, Ontario N6A 5B7, Canada
[5]Department of Chemistry, Rice University, Houston, TX 77005, USA

**Abstract**

Recently, Tao and Mo developed a new nonempirical semilocal exchange-correlation density functional. The exchange part of this functional is derived from a density matrix expansion corrected to reproduce the fourth-order gradient expansion in the slowly varying limit, while the correlation part is based on the TPSS correlation model with a modification for the low-density limit. In the present work, the Tao-Mo functional is assessed by calculations on a variety of solids and jellium surfaces. This includes 22 lattice constants and bulk moduli, 7 cohesive energies, and jellium surface exchange and correlation energies for the density parameter $r_s$ in the range from 2 to 3 bohrs. Our calculations show that this meta-generalized gradient approximation can yield consistently remarkable accuracy for the properties considered here, with mean absolute errors of 0.017 Å for lattice constants, 7.0 GPa for bulk moduli, 0.08 eV for cohesive energies, and 35 erg/cm$^2$ for surface exchange-correlation energies, substantially improving upon existing nonempirical semilocal density functionals.

1. **INTRODUCTION**

Due to the high computational efficiency and useful accuracy, the Kohn-Sham density functional theory [1] has become the most widely used method for electronic structure calculations of molecules and solids. In this theory, only the exchange-correlation energy component that accounts for all many-body effects must be approximated as a functional of the electron density. Therefore, development of accurate and widely applicable exchange-correlation energy functionals has been a primary goal of this theory.

Although many exact properties of the exchange-correlation functional have been discovered, the exact functional itself remains unknown. Approximations can be constructed by assuming some functional form that contains many parameters under the guidance of some basic properties such as uniform coordinate scaling, spin scaling, negativity of energy density, and uniform-gas limit. The parameters introduced, or part of them, are determined by a fit to experiment or highly accurate theoretical reference values for selected properties and systems. Such functionals are called empirical or semiempirical. Other functionals have been developed by imposing exact or nearly exact constraints on the density functional, so that all introduced parameters can be fixed by the imposed constraints. Approximate functionals of this type are called nonempirical. Nonempirical functionals may not be as accurate as empirical functionals for certain properties or sets of properties, but they provide a more balanced description of physically different systems such as molecules, solids, and surfaces, because the parameters determined by universal constraints are usually more easily transferable from one system to another than those determined through empirical fitting to a set of properties. This has been demonstrated by the universally good performance of the nonempirical Perdew-Burke-Ernzerhof [2] (PBE) generalized-gradient approximation (GGA) and Tao-Perdew-Staroverov-Scuseria [3] (TPSS) meta-GGA. On the other hand, empirical functionals can be highly accurate for subsets of systems and properties, pushing semilocal DFT to the accuracy limit for a particular functional form. For example, the M06L functional developed by Zhao and Truhlar [4] contains more than 30 empirical parameters. This functional shows high accuracy in quantum chemistry. However, it is relatively less accurate in condensed matter physics (e.g., the error of M06-L in lattice constants is greater than those of PBE and TPSS [5]).

Physically, the exchange-correlation energy arises from the interaction between an electron and the exchange-correlation hole surrounding the electron. The exchange-correlation hole associated with a given semilocal functional is generally unknown, but it can be found by the reverse-engineering approach. By construction, the hole is constrained to reproduce the exchange-correlation energy of the semilocal functional. There are many forms of the associated hole that can satisfy this and other constraints. Therefore, additional approximations have to be introduced in the construction of the hole.

In the development of semilocal DFT, an appealing approach is to approximate the exchange-correlation hole directly, and then derive the corresponding energy functional. Recently, Tao and Mo [6] developed a meta-generalized gradient approximation (meta-GGA) for the exchange-correlation energy. In this work, we assess the performance of the Tao-Mo (TM) meta-GGA on lattice constants, bulk moduli, cohesive energies of solids, and surface exchange and correlation energies of jellium. Our numerical tests show that this nonempirical density functional can achieve substantially improved accuracy for a variety of solids and surfaces.

## 2. Computational Method

The TM meta-GGA functional is written as [6]

$$E_{xc}\left[n_\uparrow, n_\downarrow\right] = \int d^3r\, n\epsilon_{xc}^{unif}\left(n_\uparrow, n_\downarrow\right) F_{xc}(n_\uparrow, n_\downarrow, \nabla n_\uparrow, \nabla n_\downarrow, \tau_\uparrow, \tau_\downarrow), \tag{1}$$

where $n(\mathbf{r}) = n_\uparrow(\mathbf{r}) + n_\downarrow(\mathbf{r})$ is the total electron density, $\epsilon_{xc}^{unif}\left(n_\uparrow, n_\downarrow\right)$ is the exchange-correlation energy per electron for the uniform electron gas, $F_{xc}$ is the enhancement factor, $\nabla n_\sigma(\mathbf{r})$ is the density gradient, and $\tau_\sigma(\mathbf{r}) = \frac{1}{2}\sum_i |\nabla \phi_{i\sigma}(\mathbf{r})|^2$ is the Kohn-Sham kinetic energy density of $\sigma$-spin electrons.

For a spin-unpolarized density, the exchange part of the TM meta-GGA enhancement factor is given as a weighted average of two enhancement factors: one derived from the density-matrix expansion (DME)[7] and the other from the slowly varying density correction (SC),

$$F_x = wF_x^{DME} + (1-w)F_x^{SC}. \qquad (2)$$

The DME enhancement factor is given by

$$F_x^{DME} = \frac{1}{f^2} + \frac{7}{9f^4}\left\{1 + \frac{595}{54}(2\lambda-1)^2 p - \frac{1}{\tau^{unif}}\left[\tau - 3\left(\lambda^2 - \lambda + \frac{1}{2}\right)\left(\tau - \tau^{unif} - \frac{1}{72}\frac{|\nabla n|^2}{n}\right)\right]\right\}, \qquad (3)$$

where $\tau^{unif} = 3k_F^2 n/10$ is the kinetic energy density for the uniform electron gas, $p = s^2 = (|\nabla n|/2k_F n)^2$, $k_F = (3\pi^2 n)^{1/3}$ is the Fermi wave vector, $f = \left[1 + 10(70y/27) + \beta y^2\right]^{1/10}$, $y = (2\lambda-1)^2 p$, with $\lambda = 0.6866$, and $\beta = 79.873$. In the slowly-varying density limit, the first term on the right-hand side of Eq. (3) reduces to 1, while the second term vanishes. Therefore, the DME recovers the correct uniform-gas limit, but the gradient expansion coefficients are not right. The slowly-varying density correction is needed. $F_x^{SC}$ is given by

$$F_x^{SC} = \left\{1 + 10\left[\left(\frac{10}{81} + \frac{50}{729}p\right)p + \frac{146}{2025}\tilde{q}^2 - \frac{73}{405}\tilde{q}\frac{3}{5}\left(\frac{\tau_W}{\tau}\right)\left(1 - \frac{\tau_W}{\tau}\right)\right] + 0\cdot p^2\right\}^{1/10}, \qquad (4)$$

where $\tilde{q} = 3\tau/2k_F^2 n - 9/20 - p/12$, and $\tau_W = |\nabla n|^2/8n$ is the von Weizsäcker kinetic energy density. In the slowly-varying limit, $F_x^{SC}$ reduces to the exact fourth-order gradient expansion.[6] The weight is given by

$$w = \frac{(\tau_W/\tau)^2 + 3(\tau_W/\tau)^3}{\left[1 + (\tau_W/\tau)^3\right]^2}. \tag{5}$$

For one-electron densities, $w=1$, while in the uniform-gas limit, $w=0$. In the slowly varying limit, our enhancement factor of Eq. (2) correctly reduces to $F_x^{SC}$.

The correlation part of the TM meta-GGA functional takes the same form as TPSS correlation [Eqs. (11) and (12) of Ref. 3], but replaces $C(\zeta, \xi)$ by a simpler form

$$C(\zeta, \xi) = \frac{0.1\zeta^2 + 0.32\zeta^4}{\left\{1 + \xi^2 \left[(1+\zeta)^{-4/3} + (1-\zeta)^{-4/3}\right]/2\right\}^4}, \tag{6}$$

with $\zeta = (n_\uparrow - n_\downarrow)/n$ being the relative spin polarization, and $\xi = |\nabla \zeta|/2k_F$. This modification is motivated by the fact that, in the low-density limit, correlation shows exchange-like scaling behavior, while in the high-density limit, correlation scales to a constant, indicating the significance of correlation in the low-density limit.[8] (The modification of the TPSS correlation energy is equivalent to the modification of the TPSS correlation hole, because the latter can be constructed from the former with inverse engineering approach.[9])

### 3. Results and Discussion

#### 3.1 Lattice Constants

The lattice constant of a solid at equilibrium is a basic quantity on which all other properties depend. Accurate prediction of this quantity is critical in the design of materials and devices.[10–12] Our test set of 22 bulk crystals includes main-group metals Li, K, Al, semiconductors diamond, Si, β-SiC, Ge, BP, AlP, AlAs, GaN, GaP, GaAs, ionic crystals NaCl, NaF, LiCl, LiF, MgO, MgS, and transition metals Cu, Pd,

Ag. Calculations on these solids were performed using a locally modified version [6] of the Gaussian program [13] with periodic boundary conditions (PBC). [14] Summarized in Table I are the basis sets used in the calculation of the 22 bulk solids. Gaussian-type basis set developed for atoms and molecules often contain diffuse functions. When applying Gaussian-type basis sets to solid systems, such diffuse functions should be removed for computational efficiency. For smooth convergence and reliability of results, dense $k$-point meshes were used in the evaluation of energy: $22\times22\times22$ to $20\times40\times40$ for main group metals, $10\times10\times10$ to $12\times12\times12$ for semiconductors, $10\times10\times10$ to $14\times14\times14$ for ionic crystals, and $8\times16\times16$ to $10\times18\times18$ for transition metals.

**Table I**: The Gaussian-type basis sets adopted for the atoms of the 22 bulk solids. The Strukturbericht symbols in parentheses denote the types of crystal structures: face-centered cubic (A1), body-centered cubic (A2), diamond (A4), rock salt (B1), and zinc blende (B3). The "Cartesian" configuration includes six $d$ functions. The ''pure'' configuration includes five $d$ functions.

| Solid | Basis set | | $d$ functions |
|---|---|---|---|
| Li (A2) | $4s,3p,1d$ [15] | | pure |
| K (A2) | $6s,4p,1d$ [16] | | Cartesian |
| Al (A1) | $6s,3p,1d$ [17] | | Cartesian |
| C (A4) | 6-31G* | | Cartesian |
| Si (A4) | 6-31G* | | Cartesian |
| SiC (B3) | Si: 6-31G* | C: 6-31G* | Cartesian |
| Ge (A4) | ECP-$4s,3p,2d$ [18] | | pure |
| BP (B3) | B: $4s,3p,1d$ [18] | P: $6s,5p,1d$ [18] | pure |
| AlP (B3) | Al: $6s,3p,1d$ [17] | P: 6-311G* | pure |
| AlAs (B3) | Al: $6s,3p,1d$ [17] | As: 6-311G* | pure |
| GaN (B3) | Ga: $6s,5p,2d$ [19] | N: 6-311G* | pure |
| GaP (B3) | Ga: $6s,5p,2d$ [19] | P: 6-311G* | pure |
| GaAs (B3) | Ga: $6s,5p,2d$ [19] | As: 6-311G* | pure |
| NaCl (B1) | Na: $6s,4p,1d$ [20] | Cl: 6-311G* | pure |
| NaF (B1) | Na: $6s,4p,1d$ [20] | F: 6-311G* | pure |
| LiCl (B1) | Li: $4s,3p,1d$ [20] | Cl: 6-311G* | pure |
| LiF (B1) | Li: $4s,3p,1d$ [20] | F: 6-311G* | pure |
| MgO (B1) | Mg: $4s,3p,1d$ [21] | O: $4s,3p,1d$ [21] | pure |
| MgS (B1) | Mg: $4s,3p,1d$ [21] | S: 6-311G* | pure |
| Cu (A1) | $6s,5p,2d$ [22] | | pure |
| Pd (A1) | ECP [23]-$4s,4p,2d$ | | pure |
| Ag (A1) | ECP [23]-$4s,4p,2d$ | | pure |

Listed in Table II are the equilibrium lattice constants of the 22 solids calculated with TM and literature values. Depicted in Figure 1 is the comparison of errors of LSDA, PBE, TPSS, PBEsol, revTPSS, and TM for lattice constants of these solids. The TM functional has a mean error (ME) of only 0.002 Å and is the second most balanced functional between underestimation and overestimation. The greatest reductions of error by TM compared with TPSS were achieved on K, NaF, and NaCl. Overall, TM is the most accurate functional for lattice constants, with a mean absolute error (MAE) of 0.017 Å which is the smallest of all listed. In our comparison, phonon zero-point energy (ZPE) was not removed from experiments, because this effect is small and not perfectly known. (the SCAN functional[5] also yields very accurate lattice constants.)

**Table II**: Equilibrium (0 K) lattice constants (Å) of 22 solids calculated with various functionals. The ME and MAE are in reference to experimental lattice constants. The LSDA, PBE, TPSS, PBEsol, and revTPSS values for Ge, BP, AlP, AlAs, GaN, GaP, and MgS are from Ref. 24. The other LSDA, PBE, and TPSS values are from Ref. 18, PBEsol from Ref. 25. The revTPSS results are taken from Ref. 26 except for potassium which is from Ref. 24. The experimental data for lattice constants are from Ref. 18 except those for BP, AlP, AlAs, GaN, GaP, and MgS which are from Ref. 27. The TM values are calculated self-consistently. The smallest and largest MAEs are in bold blue and red, respectively.

| Solids | Expt. | LSDA | PBE | PBEsol | TPSS | revTPSS | TM |
|---|---|---|---|---|---|---|---|
| Li | 3.477 | 3.383 | 3.453 | 3.453 | 3.475 | 3.425 | 3.445 |
| K | 5.225 | 5.093 | 5.308 | 5.232 | 5.362 | 5.325 | 5.265 |
| Al | 4.032 | 4.008 | 4.063 | 4.038 | 4.035 | 4.005 | 4.024 |
| C | 3.567 | 3.544 | 3.583 | 3.562 | 3.583 | 3.559 | 3.564 |
| Si | 5.430 | 5.426 | 5.490 | 5.442 | 5.477 | 5.437 | 5.443 |
| SiC | 4.358 | 4.351 | 4.401 | 4.381 | 4.392 | 4.358 | 4.374 |
| Ge | 5.652 | 5.624 | 5.764 | 5.679 | 5.723 | 5.680 | 5.671 |
| BP | 4.538 | 4.491 | 4.548 | 4.520 | 4.544 | 4.529 | 4.534 |
| AlP | 5.460 | 5.433 | 5.504 | 5.468 | 5.492 | 5.482 | 5.487 |
| AlAs | 5.658 | 5.631 | 5.728 | 5.676 | 5.702 | 5.682 | 5.691 |
| GaN | 4.531 | 4.457 | 4.549 | 4.499 | 4.532 | 4.518 | 4.492 |
| GaP | 5.448 | 5.392 | 5.506 | 5.439 | 5.488 | 5.460 | 5.437 |
| GaAs | 5.648 | 5.592 | 5.726 | 5.687 | 5.702 | 5.673 | 5.641 |
| NaCl | 5.595 | 5.471 | 5.698 | 5.611 | 5.696 | 5.671 | 5.618 |
| NaF | 4.609 | 4.505 | 4.700 | 4.633 | 4.706 | 4.674 | 4.626 |
| LiCl | 5.106 | 4.968 | 5.148 | 5.072 | 5.113 | 5.087 | 5.089 |
| LiF | 4.010 | 3.904 | 4.062 | 4.002 | 4.026 | 4.011 | 3.995 |
| MgO | 4.207 | 4.156 | 4.242 | 4.229 | 4.224 | 4.233 | 4.209 |
| MgS | 5.202 | 5.127 | 5.228 | 5.184 | 5.228 | 5.222 | 5.198 |
| Cu | 3.603 | 3.530 | 3.636 | 3.578 | 3.593 | 3.548 | 3.587 |
| Pd | 3.881 | 3.851 | 3.950 | 3.888 | 3.917 | 3.876 | 3.900 |
| Ag | 4.069 | 3.997 | 4.130 | 4.045 | 4.076 | 4.050 | 4.052 |
| ME | | -0.062 | 0.051 | 0.001 | 0.035 | 0.009 | 0.002 |
| MAE | | **0.062** | 0.053 | 0.018 | 0.037 | 0.028 | **0.017** |

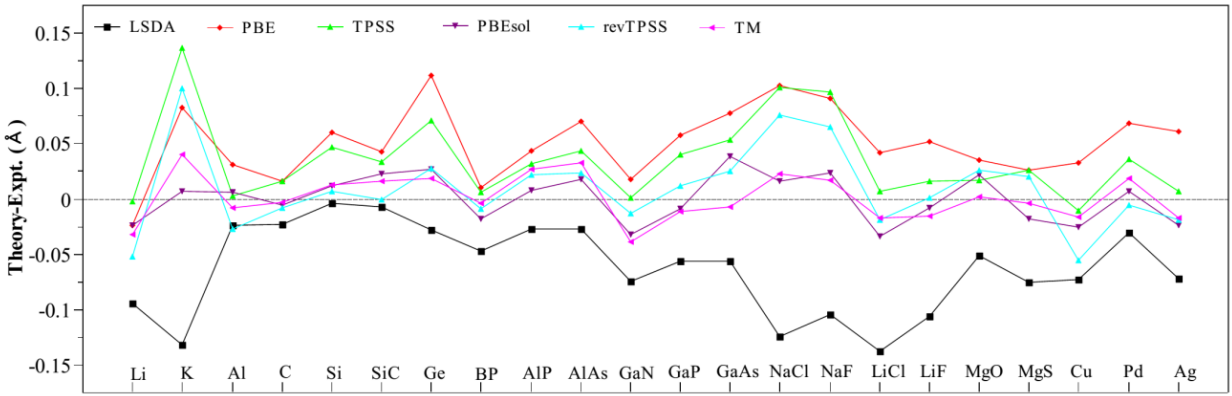

Figure 1 Deviation of lattice constants of 22 solids from experimental values at 0 K. All values are from Table II.

### 3.2 Bulk Moduli

Bulk modulus is just the curvature of total energy v.s. the volume at the minimum-energy volume. However, it presents a great challenge to DFT, in particular for transition metals.[28] It can be calculated from the equation of state. Several models have been proposed for the EOS.[29–31] These models should produce the same or nearly the same bulk modulus.

In the present work, to obtain the zero-temperature equilibrium lattice constant and bulk modulus for each crystal, calculations of the total energy was first performed on no less than 10 static lattices. The unit cells of such lattices have volumes ranging from -5% to +5% of the equilibrium cell volume. To generate the equilibrium unit cell volume and bulk modulus, the energy versus unit cell volume curve was then fitted to the stabilized jellium equation of state (SJEOS)[29,30]

$$\varepsilon(x) = \frac{a}{x^3} + \frac{b}{x^2} + \frac{c}{x} + d \ , \tag{7}$$

where $\varepsilon$ is the energy of the lattice cell, and $x$ is the volume. The equilibrium lattice volume $v_0$ and bulk modulus $B_0$ were obtained by solving

$$a = \frac{9}{2} B_0 v_0 (B_1 - 3) \ , \tag{8}$$

$$b = \frac{9}{2} B_0 v_0 (10 - 3B_1) \ , \tag{9}$$

$$c = -\frac{9}{2} B_0 v_0 (11 - 3B_1) \ . \tag{10}$$

Listed in Table III are the equilibrium bulk moduli of the 22 solids calculated with TM and other functionals. Plotted in Figure 2 are the deviations of LSDA, PBE, TPSS, PBEsol, and TM bulk moduli of the 22 crystalline solids from experimental data. The TM functional is more accurate for bulk moduli than all-purpose nonempirical functionals PBE and TPSS. The MAE of TM is 7.0 GPa, which is larger than that of PBEsol (MAE = 6.0 GPa), but smaller than those of TPSS (MAE = 8.8 GPa), PBE (MAE =

7.8 GPa), and LSDA (MAE = 12.0 GPa). In comparison with PBEsol, TM has the best accuracy for diamond, while the error is comparatively large for Ag, MgO, and Cu.

**Table III**: Equilibrium bulk moduli (GPa) of the 22 solids calculated at 0 K. The LSDA, PBE, and TPSS values are from Ref. 20. For BP, AlP, AlAs, GaN, GaP, and MgS, the LSDA and PBE values are from Ref. 26, the PBEsol values are from Ref. 27. The experimental values of bulk moduli for the 22 solids are from the following references: Li,[34] K,[35] Al,[36] C[37], Si,[38] SiC,[39] Ge,[38] BP,[40] AlP,[41] AlAs,[41] GaN,[42] GaP,[41] GaAs,[38] NaCl,[43] NaF,[43] LiCl,[43] LiF,[44] MgO,[45] MgS,[46] Cu,[47] Pd,[48] and Ag.[49] The smallest and largest MAEs are in bold blue and red, respectively.

| Solids | LSDA | PBE | TPSS | PBEsol | TM | Expt. |
|---|---|---|---|---|---|---|
| Li | 14.7 | 13.7 | 13.2 | 13.8 | 13.7 | 13 |
| K | 4.6 | 3.8 | 3.6 | 3.7 | 4.0 | 3.7 |
| Al | 82.5 | 76.8 | 85.2 | 82.6 | 88.6 | 79.4 |
| C | 458 | 426 | 421 | 450.0 | 442.4 | 443 |
| Si | 95.6 | 89 | 91.9 | 94.2 | 97.1 | 99.2 |
| SiC | 225 | 209 | 213 | 218.0 | 220.0 | 225 |
| Ge | 75.9 | 63.0 | 66.4 | 68.1 | 72.5 | 75.8 |
| BP | 176 | 162 | | | 171.5 | 173 |
| AlP | 89.9 | 82.6 | | | 89.3 | 86 |
| AlAs | 75.5 | 67.0 | | | 75.2 | 82 |
| GaN | 204 | 173 | | | 207.1 | 190 |
| GaP | 90.6 | 77.0 | | | 89.2 | 88 |
| GaAs | 81.3 | 68.1 | 70.1 | 69.1 | 78.6 | 75.6 |
| NaCl | 32.5 | 23.9 | 23 | 25.8 | 26.9 | 26.6 |
| NaF | 63.3 | 47.7 | 44 | 48.6 | 52.5 | 51.4 |
| LiCl | 42 | 32.9 | 34.3 | 35.2 | 36.2 | 35.4 |
| LiF | 87.5 | 65.9 | 67.2 | 73.1 | 74.4 | 69.8 |
| MgO | 183 | 162 | 169 | 157.0 | 174.5 | 165 |
| MgS | 84.0 | 74.4 | | | 79.8 | 78.9 |
| Cu | 192 | 153 | 173 | 166.0 | 180.2 | 142 |
| Pd | 240 | 180 | 203 | 205.0 | 210.7 | 195 |
| Ag | 153 | 107 | 129 | 119.0 | 138.4 | 109 |
| ME | 11.1 | -6.8 | -0.1 | 1.3 | 5.3 | |
| MAE | **12.0** | 7.8 | 8.8 | **6.0** | 7.0 | |
| MARE | 0.1 | 0.1 | 0.1 | 0.1 | 0.1 | |

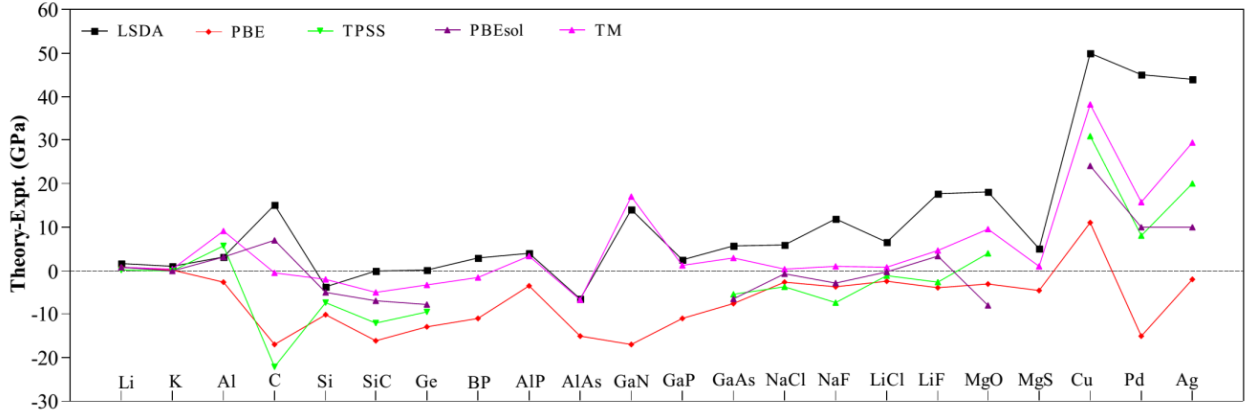

Figure 2 Deviation of 0 K bulk moduli of 22 solids from experimental values. All values are from Table III.

### 3.3 Cohesive Energies

Cohesive energy is the difference between the total energy of a bulk solid and the constituent neutral atoms. It is the condensed-matter analog of the molecular atomization energy and a measure of the interatomic bond strength. To compute the cohesive energy for each of the 7 solids, the total energy of the equilibrium lattice was first divided by the number of atoms to get the total energy per atom. This electronic total energy per atom was then corrected by adding the phonon ZPE to account for the zero-point motion. The phonon ZPE per atom is approximated via [29]

$$\varepsilon_{\text{ZPE}} = \frac{9}{8} k_B \Theta_D , \qquad (11)$$

where $k_B$ is the Boltzmann constant and $\Theta_D$ is the Debye temperature of the solid. In the present work, we adopted the following Debye temperatures: C 2230K,[50] Si 645K,[50] SiC 1232K,[51] NaCl 321K,[50] NaF 492K,[50] LiCl 422K,[50] and LiF 732K.[50] The ZPE-corrected energy per atom was then subtracted from the ground-state energy of isolated atoms with spin freedom to obtain the cohesive energy. Among the 6 atoms (C, Si, Na, Li, Cl, F) comprising the 7 solids, the atoms Li and Na involve diffuse functions in their molecular basis sets. These diffuse basis functions were excluded in the calculations of lattice constants and bulk moduli of ionic solids that contain Li or Na, but used for calculating the ground-state energies of the isolated Li and Na atoms (i.e., the full molecular basis set 6-311G* was employed for the isolated Li and Na atoms). Applying two different basis sets to the solid and the corresponding isolated atoms provides reasonable cohesive energies especially in the case of these ionic solids, because cations are compact and their electrons are less likely to appear in the far regions described by diffuse functions, therefore decreasing the need of diffuse functions in the solid-state calculation. Listed in Table IV are the cohesive energies (eV/atom) of 7 solids. Figure 3 compares the performance of the LSDA, PBE, TPSS, PBEsol, revTPSS, and TM functionals for cohesive energies of these 7 solids. TM has a mean absolute relative error (MARE) of only 1.9%, nearly half that of the meta-GGA TPSS (MARE=3.7%) and smaller than such errors for revTPSS (MARE=3.4%), PBE (MARE=2.5%), PBEsol (MARE=4.2%), and LSDA (MARE=13.4%). This is in sharp contrast with atomization energies of molecular systems [6] in which TM is less accurate than TPSS for the 148 G2 molecules and moderately more accurate than TPSS for the AE6 test set.

**Table IV**: Cohesive energies (eV/atom) of 7 solids. The LSDA, PBE, and TPSS values are from Ref. 20, PBEsol from Ref. 33, and revTPSS from Ref. 52. The TM values are calculated self-consistently and corrected for zero-point vibrations. The smallest and largest MAEs are in bold blue and red, respectively.

| Solid | LSDA | PBE | TPSS | PBEsol | revTPSS | TM | Expt. |
|---|---|---|---|---|---|---|---|
| C | 8.83 | 7.62 | 7.12 | 8.05 | 7.31 | 7.48 | 7.37 |
| Si | 5.26 | 4.50 | 4.36 | 4.87 | 4.50 | 4.61 | 4.62 |
| SiC | 7.25 | 6.25 | 6.02 | 6.75 | 6.26 | 6.29 | 6.37 |
| NaCl | 3.58 | 3.16 | 3.18 | 3.20 | 3.14 | 3.19 | 3.31 |
| NaF | 4.50 | 3.96 | 3.87 | 3.99 | 3.74 | 3.88 | 3.93 |
| LiCl | 3.88 | 3.41 | 3.41 | 3.49 | 3.39 | 3.42 | 3.55 |
| LiF | 5.02 | 4.42 | 4.32 | 4.49 | 4.23 | 4.34 | 4.40 |
| ME | 0.68 | -0.03 | -0.18 | 0.18 | -0.14 | -0.05 | |
| MAE | **0.68** | 0.12 | 0.18 | 0.23 | 0.14 | **0.08** | |
| MARE | 13.4 | 2.5 | 3.7 | 4.2 | 3.4 | 1.9 | |

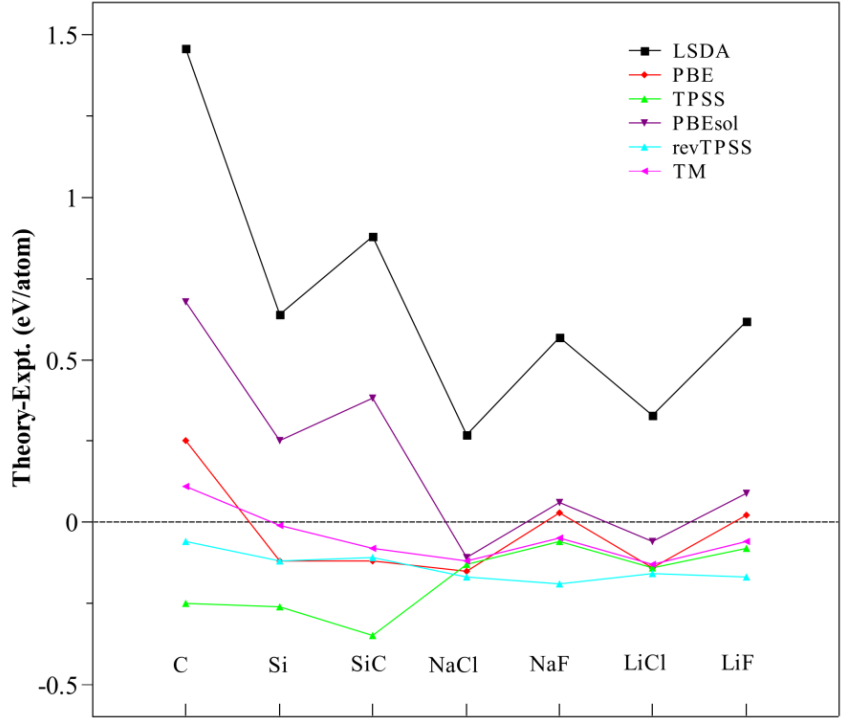

Figure 3 Deviation of cohesive energies of 7 solids from experimental values. All values are from Table IV.

### 3.4 Surface Exchange and Correlation Energies

Jellium, a homogeneous electron gas with a positive uniform background charge, is a realistic model for simple metals. The electron density of jellium is uniform within the bulk, while near the surface it is rapidly varying and exponentially decaying in vacuum. The surface energy per unit area $\sigma$ is defined as the energy difference between the bulk and surface. The exchange-correlation contribution can be calculated from

$$\sigma_{xc} = \int_{-\infty}^{\infty} n(z)[\epsilon_{xc}(z) - \epsilon_{xc}(-\infty)]dz, \tag{12}$$

From this equation, we can see that, in order to have an accurate description of the surface energy, a density functional must be correct in both the slowly varying and rapidly varying regimes, in particular the slowly varying gradient expansion.

For this model system, several *ab initio* calculations of the surface energy are available in the literature, including the random-phase approximation [(RPA)[53] and quantum Monte Carlo (QMC)[54]]. These calculations agree well with each other and with time-dependent DFT.[53] Since QMC values have some uncertainty, we compare all the DFT values to the RPA calculation in the high-density regime from $r_s = 2$ bohrs to $r_s = 3$ bohrs, in which the RPA is reliable. The results displayed in Table V show that the surface exchange energy of TM exchange functional is in excellent agreement with the exact values,[55] dramatically improving upon the LSDA, PBE, and TPSS. The TM functional has an MAE of only 10 erg/cm$^2$, much smaller than the MAEs of LSDA (MAE=284 erg/cm$^2$), PBE (MAE=125 erg/cm$^2$), and TPSS (MAE=51 erg/cm$^2$). This excellent performance of the TM functional largely benefits from the recovery of the correct fourth-order gradient expansion in the slowly varying limit and good asymptotic behavior. In total, TM yields much better surface exchange-correlation energy (MAE=35 erg/cm$^2$) than LSDA (MAE=77 erg/cm$^2$), PBE (MAE=133 erg/cm$^2$), and TPSS (MAE=60 erg/cm$^2$).

**Table V**: Jellium surface exchange energies $\sigma_x$ and surface exchange-correlation energies $\sigma_{xc}$ (in erg/cm$^2$). The reference values are taken from the RPA calculation.[55] The LSDA, PBE, and TPSS values are taken from Ref. 20. The smallest and largest MAEs are in bold blue and red, respectively.

| $r_s$ (bohr) | Exchange | | | | | Exchange-correlation | | | | |
| --- | --- | --- | --- | --- | --- | --- | --- | --- | --- | --- |
| | LSDA | PBE | TPSS | TM | RPA | LSDA | PBE | TPSS | TM | RPA |
| 2.00 | 3037 | 2438 | 2553 | 2641 | 2624 | 3354 | 3265 | 3380 | 3515 | 3467 |
| 2.07 | 2674 | 2127 | 2231 | 2312 | 2296 | 2961 | 2881 | 2985 | 3109 | 3064 |
| 2.30 | 1809 | 1395 | 1469 | 1531 | 1521 | 2019 | 1962 | 2035 | 2132 | 2098 |
| 2.66 | 1051 | 770 | 817 | 860 | 854 | 1188 | 1152 | 1198 | 1267 | 1240 |
| 3.00 | 669 | 468 | 497 | 528 | 526 | 764 | 743 | 772 | 823 | 801 |
| ME | 284 | -125 | -51 | 10 | | -77 | -133 | -60 | 35 | |
| MAE | **284** | 125 | 51 | **10** | | 77 | **133** | 60 | **35** | |

## 4. CONCLUSIONS

In summary, we have evaluated the performance of the TM meta-GGA for solids and solid surfaces. Our results show that this functional is consistently accurate for lattice constants, bulk moduli, cohesive energies, and surface exchange-correlation energies, substantially improving upon the nonempirical density functionals proposed in recent years. Our evaluation suggests that this novel meta-GGA functional is a good choice for electronic structure calculations in condensed matter physics.

**Acknowledgments**

We thank Professor Guocai Tian for many helpful discussions and valuable suggestions. This work was supported by NSF under Grant no. CHE-1640584. Computational support was provided by Temple University. VNS was supported by the Natural Sciences and Engineering Research Council of Canada (NSERC). The work at Rice University was supported by the U.S. Department of Energy, Office of Basic Energy Sciences, Computational and Theoretical Chemistry Program under Award No. DE-FG02-09ER16053. G.E.S. is a Welch Foundation Chair (C-0036).

* Corresponding author. jianmin.tao@temple.edu